\documentclass[prl,superscriptaddress,amsmath,amsfonts,twocolumn]{revtex4-1}
\usepackage{graphicx}
\usepackage{epsfig}
\usepackage{amssymb}

\usepackage[hyperindex,breaklinks]{hyperref}

\usepackage{titlesec}
\titlespacing*{\subsection}{0pt}{.65\baselineskip}{.65\baselineskip}

\usepackage{placeins}

\newcommand {\tpsi}{\tilde{\psi}}

\newcommand {\tphi}{\tilde{\phi}}

\newcommand{\PT}{{\cal PT}}

\newcommand{\tA}{\tilde{A}}
\newcommand{\tQ}{\tilde{Q}}

\newcommand{\tq}{\tilde{q}}
\newcommand{\tg}{\tilde{g}}

\newcommand{\bF}{{\bf F}}
\newcommand{\bE}{{\bf E}}
\newcommand{\bH}{{\bf H}}

\DeclareMathOperator{\sech}{sech}
\usepackage[normalem]{ulem}
\hypersetup{backref,colorlinks=true,linkcolor=blue,citecolor=blue,urlcolor=blue,linktocpage=true,breaklinks=true}

\begin{document}

\title{ {\fontfamily{ptm}\selectfont  Waveguides with Absorbing Boundaries: Nonlinearity Controlled by an Exceptional Point and Solitons}}

\author{{\bf  Bikashkali Midya}}
\email{bmidya@ist.ac.at}
\affiliation{Institute of Science and Technology Austria, Am Campus 1, 3400 Klosterneuburg, Austria }

\author{{\bf Vladimir V. Konotop}}
\email{vvkonotop@fc.ul.pt}
\affiliation{Centro de F\'isica Te\'orica e Computacional and Departamento de F\'isica, Faculdade de Ci\^encias, Universidade de Lisboa, Campo Grande 2, Edif\'icio C8, Lisboa 1749-016, Portugal}

%\date{\today}

\begin{abstract} 
{ We reveal the existence of continuous families of guided single-mode solitons in planar waveguides with weakly nonlinear active core and absorbing boundaries. Stable propagation of TE and TM-polarized solitons   is accompanied by attenuation of all other modes, i.e., the waveguide features properties of conservative and dissipative systems.  If the linear spectrum of the waveguide possesses exceptional points, which occurs in the case of TM polarization, an originally focusing (defocusing) material nonlinearity may become effectively defocusing (focusing). This occurs due to the geometric phase of the carried eigenmode when the surface impedance encircles the exceptional point. In its turn the change of the  effective nonlinearity ensures the existence of dark (bright) solitons in spite of focusing (defocusing) Kerr nonlinearity of the core. The existence of an exceptional point can also result in anomalous enhancement of the effective nonlinearity. In terms of practical applications, the nonlinearity of the reported waveguide can be manipulated by controlling the properties of the absorbing cladding.}
\end{abstract}

\maketitle

 Localized solutions of one-dimensional (1D) nonlinear conservative guiding systems are known to belong to continuous families characterized by the dependence of the mode intensity on the propagation constant (or frequency, or chemical potential, depending on the physical system). In a broad context such modes are called solitons~\cite{Dodd}. In contrast, localized solutions of nonlinear dissipative 1D systems are isolated points in the functional space. When stable, they are attractors, whose characteristics depend on the system parameters, and are cited as dissipative solitons~\cite{Akhmediev}. While solitons emerge from the balance between the nonlinearity and dispersion, dissipative solitons require also the balance between gain and loss~\cite{dissipative}. There are two known exceptions of this rule. The first one is the parity-time ($\PT$) symmetric~\cite{Bender} systems where the symmetry of the real and imaginary parts of the complex potential ensures the balance between gain and loss without need of additional constraints. Such modes were found for the optical systems governed by the nonlinear Schr\"odinger (NLS) equation with  $\PT$-symmetric potentials~\cite{Christodoulides}, and  are widely investigated in numerous applications~\cite{spec_issue,review}. The second type of nonconservative systems supporting families of nonlinear modes is a NLS equation with Wadati potentials~\cite{Wadati}, whose conservative part  has a specific relation to the gain and loss landscapes~\cite{Tsoy,ZK_OL,review,YN}. This was numerically found in~\cite{Tsoy}, explained in~\cite{ZK_OL}, and in~\cite{YN} it was argued that  no other potentials admit  soliton families.  Conceptually,  the coexistence of conservative and dissipative regimes, is also known for dynamical systems described by time-reversible Hamiltonians~\cite{PoOpBa}.

The situation can be different, if a system is not strictly 1D and there exist additional governing parameters. In this Letter we report a wide class of nonlinear waveguides with gain at the core and  {loss}  {at the cladding}, which nevertheless support propagation of continuous families of quasi-1D solitons. The underlying physical idea is a setting where the gain and {loss} are controlled by different mechanisms affecting the carrier wave itself rather than its envelope.  Such a waveguide features properties of an open system: the parameters of solutions are determined by the balance between gain and loss.  On the other hand, it supports continuous families of solitons, i.e. obeys properties of a conservative system. Moreover,   the type of the nonlinearity of such a system is controlled by the gain and  {loss}.   A waveguide with a defocusing (focusing) Kerr dielectric in the core can manifest effective focusing (defocusing) nonlinearity felt by a propagating beam. This effect occurs only if there exists an exceptional point (EP) in the linear spectrum of the waveguide, i.e., the point where two (or more) eigenvalues and eigenfunctions  coalesce~\cite{Kato}, and represents a manifestation of the topological geometric phase which is acquired by eigenmodes when encircling the EP in the parameter space~\cite{geom_phase,Rotter}.

The relevance of EPs in physics was recognized more than a century ago. The Voigt wave~\cite{Voigt}, which is the coalescence of two plane waves propagating in absorbing crystals having singular axes, exists at the EP of the dielectric tensor~\cite{general-Voigt,MacLak}. Recently, the importance of EPs was demonstrated in experiments with microwave cavities~\cite{microcav}, laser systems~\cite{EP_laser}, waveguides~\cite{waveguide_EP}, multilayered structures~\cite{multilaiered}, and optomechanical systems~\cite{optomechanical}, to mention a few. 

If a system is nonlinear, an EP in the spectrum of its linear limit still influences the propagation~\cite{EP_laser,Jianke,ZezKon,Bludov}. However, usually it is not considered as a factor affecting the nonlinear properties of the system itself.  In this Letter we show how an EP can modify the effective nonlinearity  of the medium, in particular, changing its type. 
 
Consider a planar waveguide consisting of an active medium characterized by the dielectric constant $\epsilon = \epsilon_r + i\epsilon_i$, with $\epsilon_{r}>0$ and $\epsilon_i<0$, which is bounded by two parallel absorbing layers located {at} the planes $y=\pm \ell$.  The medium obeys Kerr nonlinearity and is allowed to have nonlinear absorption (nonlinear gain is treated similarly); i.e., it is described by the Kerr coefficient $\chi_{\!_{NL}}=|\chi_{\!_{NL}}|e^{i\varphi_\chi}$, where  $\varphi_\chi \in [0,\pi]$ characterizes both the type of the nonlinearity and the relative strength of the nonlinear absorption. The medium is focusing if $\varphi_\chi \in [0,\pi/2)$ and defocusing if $\varphi_\chi \in (\pi/2,\pi]$.  At $\varphi_\chi=\pi/2$ the nonlinearity is purely absorbing.  

Let $\bF$ be a monochromatic field,  either electric $\bE$ or magnetic $\bH$  for TE or TM polarizations, respectively, which is polarized along the $\hat{\bf{x}}$ direction and propagates along the $\hat{\bf{z}}$ direction.~It solves the Helmholtz equation  
\begin{eqnarray}
\label{Helmholtz}
 \nabla^2 \bF + \ell^2 k_0^2\epsilon ~\bF+ \chi |\bF|^2 \bF =0
\end{eqnarray}
We use the dimensionless variables measuring the coordinates  in the units of $\ell$,  $k_0= \omega/c$, $\omega$ being the frequency, and $\chi=4\pi\chi_{\!_{NL}}(\ell k_0)^2$ being the {\em material nonlinearity}.   To simplify  the model, we choose a waveguide whose linear properties were previously studied~\cite{MK}. Namely, we consider that each of the absorbing boundaries is characterized by an impedance $\eta$ and that the {fields satisfy} the impedance boundary conditions which can be written as~\cite{Senior} ${\bf n \times E}=\eta{\bf H}$, where ${\bf n}$ is the normal to the cladding outwards the waveguide core. This choice is justified when the modulus of the effective dielectric permittivity of cladding is large, $|\epsilon_{\rm{clad}}| \gg 1$.    
  
Because of the active filling, even in the presence of absorbing boundaries one can find waveguide parameters assuring simultaneous guidance of one mode, weak attenuation of a few modes, and strong absorption of all other modes. This selectivity stems from different conditions of balance between gain and loss for modes having different transverse distributions.  If a solution of the linear problem, i.e., of Eq.~(\ref{Helmholtz}) at $\chi=0$, is chosen in the form of a superposition of the guided and weakly decaying modes, an expected effect at weak material nonlinearity,  $|\sqrt{\chi} ~ \bF|^2 \ll 1$, is the existence of solitons. 

\begin{figure*}[t!]
\includegraphics[width=1.43\columnwidth]{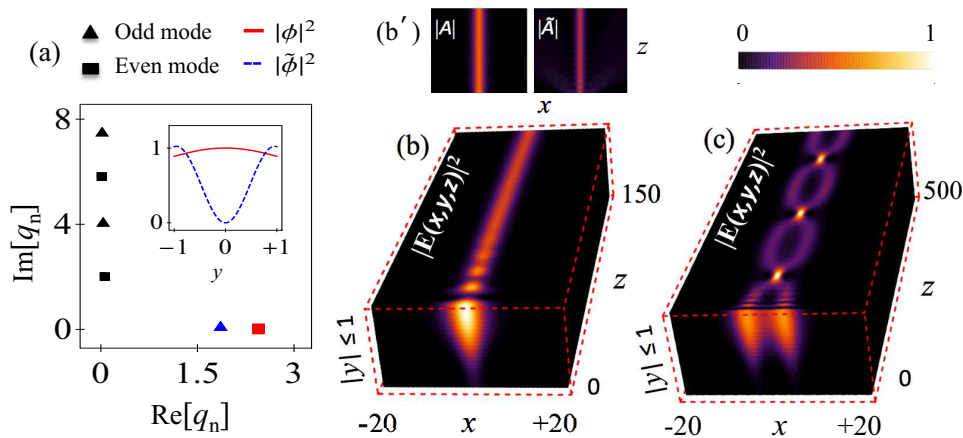}
\caption{ (a) Real part {\it vs} imaginary part of the propagation constants 	for the waveguide with~$\eta^{TE} =  -1.25 - 3.16i$,  $\epsilon = 1.5 - 0.062i$,  $|\chi_{\!_{NL}}| = 0.025$ and $k_0\ell = 2 \pi/3.1$. Squares and triangles represent the $\cos(Q y)-$ and $\sin(Q y)-$modes, respectively. The guided and weakest decaying modes are indicated by red and blue color, respectively; their transverse profiles are shown in the inset. (b) Dynamics of the single bright TE-soliton input $A = 2\tilde{A} = 0.5 \sech( \sqrt{g/8} x)$  with the amplitude profiles shown in (b$^\prime$) (see also \cite{suppl}).  (c) Dynamics of the in-phase two-soliton input: $A = 2 \tilde{A} = 0.5 [\sech(\sqrt{g/8} (x+5)) + \sech( \sqrt{g/8} (x-5))]$.
		 The effective nonlinearities are  $g = 1.24$, $\tg = 1.02 + 0.02i$, $g_1 = 1.34 + 0.07i$, $\tg_1 = 2.4 + 0.002i$. Simulations were carried out on the window
		 $-200<x<200$ and $0<z<1000$. The inputs were perturbed by noise of order 5\% of the amplitude.}
	\label{fig:one} 
\end{figure*}

\begin{figure*}[t!]
	\includegraphics[width=1.35\columnwidth,height=11.75cm]{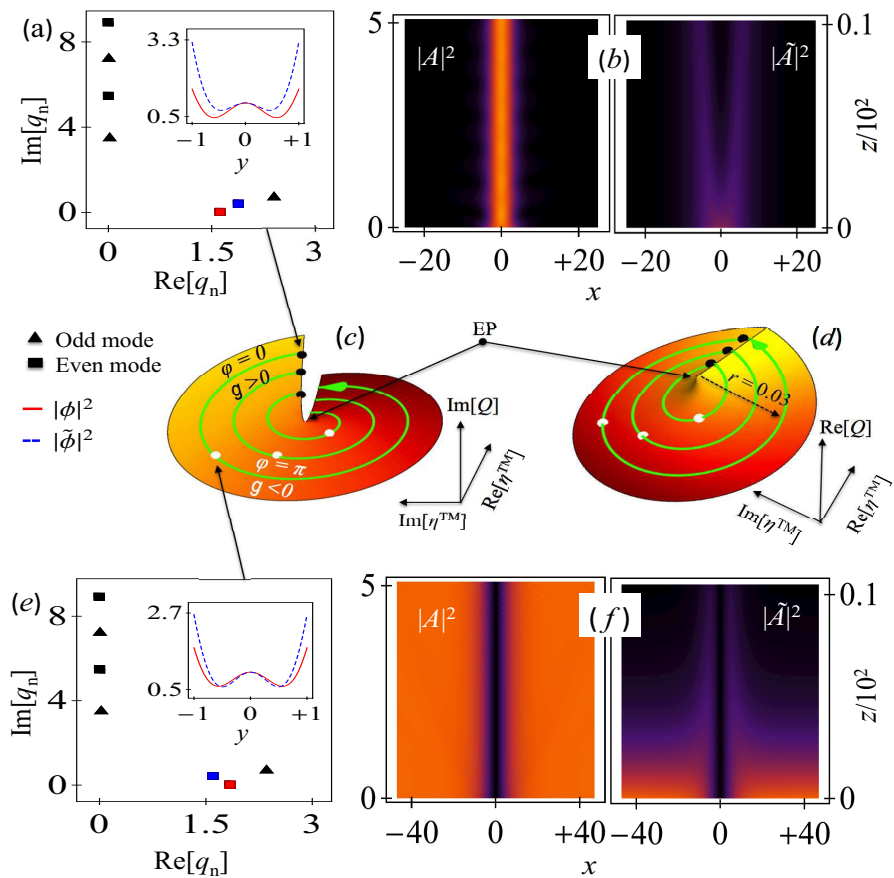}
	\caption{Central panels: Imaginary (c) and real (d) parts of the eigenvalue $Q$ corresponding to a TM-mode in the vicinity of EP, $\eta^{EP}= 1.6506 + 2.05998i$, {\it vs} real and imaginary parts of the impedance, for $0< r<0.04$ and $0\le \Theta \le 2 \pi$. The mode is chosen to be guided at two values of the total phase $\varphi$: at black discs  $\Theta = 0$, $\varphi = 0$ and $g >0$, while white discs represent the same points after change of the impedance resulting in $\varphi = \pi$ and $g <0$.  {The  positions of the phase changes are shown at $(r, \Theta) = (0.01, 1.25\pi), (0.02, 0.91\pi)$ and $(0.03, 0.722\pi)$, when $\chi= 0.12+1.29i, 0.61+1.14i$, and $0.87 + 0.95 i$, respectively.} Different $\chi$ for different radii ensure that $g$ is real at both white and black points.  Upper panels: (a) Propagation constants, and profiles of the guided and weakly decaying TM modes (in the inset), and (b) dynamics of a bright TM-soliton with $A = 2 \tilde{A} = 0.5 \sech(\sqrt{g/8} x)$ at the input, for $g = 1.42$ at $r = 0.03$ [black discs in (c) and (d)]  observed for $\eta^{TM} = 1.681 + 2.060i$, $\epsilon = 1.5  - 0.986i$, $q=1.63$, and $\tq=  1.897 + 0.386 i$. Lower panels:  (e) {Propagation constants,} and the guided and weakly decaying modes (in the inset) and  (f) dynamics of a dark TM-soliton excited by the $A = 2 \tilde{A} = 0.5 \tanh( \sqrt{|g|/8} x)$ input for $g = -0.073$ at $r = 0.03$ [white discs  in (c) and (d)] obtained  for $\eta^{TM} = 1.631 + 2.083 i$, $\epsilon = 1.5 - 0.999 i$, $ q= 1.85$, and $\tq= 1.609 + 0.409i$.  In all panels $k_0\ell=2\pi/3.1$.   Simulations were carried out on the window $-200<x<200$ and $0<z<1000$. Inputs were perturbed by noise of order 5\% of the amplitude.}
	\label{fig:two} 
\end{figure*}

 We show this for a waveguide with one guided and one weakly absorbed mode [see Fig.~\ref{fig:one} (a), and Figs.~\ref{fig:two}(a) and \ref{fig:two}(e) below]. The propagating modes are searched in the form $F\sim e^{iqz}\phi(y)$, where $q$ is the propagation constant. The   transverse profile of the mode $\phi(y)$ is determined from the non-Hermitian Sturm-Liouville  eigenvalue problem $\phi_{yy}= - Q^2\phi$ subject to the impedance (alias Robin) boundary conditions: 
$\phi^{TE}(\pm 1) = \pm\eta^{TE} \phi_y^{TE}(\pm 1)$ with $\eta^{TE} = \eta c /(i \omega \ell)$ for TE modes, and 
$\phi_y^{TM}(\pm 1) = \pm\eta^{TM} \phi^{TM}(\pm 1)$  with   {$\eta^{TM} = i\omega\epsilon\ell \eta/c$} for TM modes.  
Since the dielectric permittivity and surface impedance are complex, the eigenvalue {$Q = Q' + i Q''$} is complex, as well.  Nevertheless, the propagation constant of the guided mode $q= (\ell^2 k_0^2 \epsilon -Q^2)^{1/2}$ is real,  if
\begin{equation}
\label{cond_epsilon}
\epsilon_r > [(Q')^2 -  (Q'')^2]/ \ell^2 k_0^2 \quad\mbox{and}\quad \epsilon_i = 2 Q' Q'' / \ell^2 k_0^2.
\end{equation}
All other modes (marked by the subindex $n$) are absorbed if the condition $Q' Q'' > Q_n' Q_n''$ is verified. To distinguish the weakest absorbing mode, below we use a tilde, i.e., $\tphi$, $\tQ$ and $\tq$. For such a mode $\tq= (\ell^2 k_0^2 \epsilon -\tQ^2)^{1/2} =\tq'+i\tq''$,  where $\tq'' > 0$, and $|\tq''|\ll |\tq'|$.

We start with a waveguide whose linear spectrum does not feature EPs. Let $\psi(y)$ be an eigenfunction of the Sturm-Liouville problem adjoint to the above one for $\phi(y)$.  The states $\{\phi,\tphi,\phi_2,\phi_3,...\}$ and $\{\psi,\tpsi,\psi_2,\psi_3,...\}$ constitute a complete biorthogonal basis~\cite{suppl},  which is endowed with the scalar product $\langle\psi,\phi\rangle= \int_{-1}^{1} \psi^*(y)\phi(y)dy$. In particular, $\langle\psi,\tphi\rangle=\langle\tpsi,\phi\rangle=0$.  The eigenfunctions  $\psi = \phi^*$ and $\tpsi = \tphi^*$ correspond to the eigenvalues $Q^*$ and $\tQ^*$. 

Next, we look for a solution of Eq.~(\ref{Helmholtz}) in the form $F \approx A(x, z)\phi(y)e^{iqz} + \tA(x,z)\tphi(y)e^{i\tq z},$ where  $A$ and $\tA$ are the slowly varying amplitudes of the modes. Performing the  multiple-scale analysis~\cite{suppl}, we obtain coupled NLS equations  
\begin{eqnarray}
\label{NLS1}
2iq A_z+A_{xx}+(g |A|^2+ g_{1}e^{-2\tq''z}|\tA|^2)A=0,
\\
\label{NLS2}
2i\tq'\tA_z+\tA_{xx}+(\tg_{1} |A|^2+ \tg e^{-2\tq''z}|\tA|^2)\tA=0,
\end{eqnarray}
where the complex nonlinear coefficients describing {\em effective} self-phase and   cross-phase modulations are   
\begin{equation}
\label{g} 
\begin{array}{ll}
g=\chi\langle\psi,|\phi|^2\phi\rangle/\langle\psi,\phi\rangle, ~&
g_1=2\chi{\langle\psi,|\tphi|^2\phi\rangle}/\langle\psi,\phi\rangle,
\\
\tg=\chi\langle\tpsi,|\tphi|^2\tphi\rangle/\langle\tpsi,\tphi\rangle, & \tg_1=2\chi\langle\tpsi,|\phi|^2\tphi\rangle/\langle\tpsi,\tphi\rangle.
\end{array}
\end{equation}
Since $\tphi$ is the  most weakly decaying mode, at the propagation distance  $z\gtrsim 1/\tq''$ the effect of all decaying modes on the guided one, i.e. on $A$, can be neglected. After that distance the guided mode  {$\phi$} is the only one, which  propagates   {with the amplitude} governed by the NLS Eq.~(\ref{NLS1}) with $\tA=0$. This however, does not guarantee yet undistorted propagation because generally speaking $g$ is complex. In order to obtain the conservative NLS equation, which is exactly integrable and thus possesses soliton (as well as multi-soliton) solutions~\cite{Dodd}, we additionally have to require $g$ to be real. To this end we define the argument $\varphi_{g}=\arg\left(\langle\psi,|\phi|^2\phi\rangle/\langle\psi,\phi\rangle\right) \in [-\pi, \pi]$. Then Eq.~(\ref{NLS1}) with $\tA=0$ becomes the  conservative NLS equation only if  either $\varphi_\chi=-\varphi_{g}$ at $\varphi_{g}\in [-\pi, 0]$ or $\varphi_\chi=\pi-\varphi_{g}$ at $\varphi_{g}\in [0, \pi]$ is satisfied. This  leads us to several interesting conclusions.
	
First, if the total phase $\varphi=\varphi_\chi+\varphi_g$ of the effective nonlinearity $g$ is either $0$ or $\pi$ {\em the absorbing boundaries may support propagation of a single-mode soliton}, by attenuating all other modes. Second, since solitons of the NLS equation constitute two-parametric families~\cite{Dodd}, they are characterized by amplitudes and by velocities, {\em the waveguide   supports continuous families of the propagating spatially localized beams}, i.e. behaves in this respect like a conservative system. Third, it is possible to {choose} the waveguide parameters such that the effective nonlinearity $g$ for a guided mode has  opposite signs compared to the sign of the {physical nonlinearity} $\chi$  of the waveguide core. Specifically, for the nonlinear absorption considered here we have  
	\begin{eqnarray}
\label{g_chi}
\begin{array}{ll}
  g > 0\quad \mbox{ and}\quad\mbox{ Re$\,\chi <0$} \quad \mbox{if 	 $ \varphi_{g} \in  [-\pi,-\frac{\pi}{2}]$},
  	\\
  	g < 0\quad \mbox{ and}\quad\mbox{ Re$\,\chi >0$}\quad \mbox{if 	 $ \varphi_{g} \in  [\frac{\pi}{2},\pi]$}.
  	\end{array}
	\end{eqnarray}
  Thus {\em the combined effect of the (linear) boundary absorption with (linear) gain of the active media may result in the change of the type of the effective nonlinearity}. Then focusing (defocusing) material nonlinearity of the core becomes effectively defocusing (focusing). Consequently, this may result in the guidance of bright (dark) solitons even in the defocusing (focusing) material nonlinearity of the dielectric filling. 
	
Now we turn to examples of waveguide architecture supporting soliton propagation.  The consideration will be restricted to nonmagnetic claddings characterized by the positive dielectric constant, $\epsilon_{\rm clad}>0$, which corresponds to natural materials (see~Ref.~\cite{MK} for  examples).  This last requirement imposes conditions on the effective impedances~\cite{MK}: $\mbox{Re}~\eta^{TE},~\mbox{Im}~\eta^{TE}~<~0$,~and~$\mbox{Re}~\eta^{TM},~\mbox{Im}~\eta^{TM}~>~0$. The desired parameters can be achieved by adjusting the wave number $k_0$, the waveguide width  $\ell$, the nonlinear susceptibility, $\chi_{\!_{NL}}$, and the impedance $\eta$. The dielectric permittivity is not considered as an adjustable parameter, because the condition of mode guiding [Eq.~\ref{cond_epsilon}] fixes it as soon as the respective impedance is chosen.   
 
{\it TE soliton --}   For TE-polarized modes, we have found that gain and loss do not change  the sign of the effective nonlinearity in the whole domain of the explored parameters. Figure~\ref{fig:one}(a) illustrates propagation constants for the parameter choice ensuring the existence of one guided (red square) and one weakly absorbed (blue triangle) mode. All other modes (the lowest ones are shown in black) are strongly absorbed. The transverse profile of the fundamental (guided) mode is given by  $\phi(y) = \cos(Q y)$ with $Q \approx 0.453 - 0.279i$, and the propagation constant is~$q \approx 2.457$.  We also compute~$\varphi_{g} \approx - 0.0027$ and, hence, one has to  choose  $\chi = 1.3 + 0.0035i$, in order to ensure real $g$.  The transverse profile of the weakly decaying mode is described by $\tphi(y) = \sin(\tQ y)$, where~$\tQ \approx 1.65 - 0.158i$, and $\tq \approx 1.86 + 0.073i$.  

The direct numerical simulations of Eqs.~\eqref{NLS1} and \eqref{NLS2}, are
shown in  Figs.~\ref{fig:one} (b)--(c). In Fig.~\ref{fig:one} (b) we observe stable propagation of a single soliton carried by the fundamental mode after the second mode is absorbed by the structure (this is clearly visible on the 3D figure), while the upper inset shows nondecaying evolution of the soliton $A$, and an accompanying mode soliton $\tA$ (after the distance $z\gtrsim1/\tq''$ the soliton is not affected by the decaying mode, because Eqs.~(\ref{NLS1}) and (\ref{NLS2}) become effectively decoupled). In Fig.~\ref{fig:two} (c) we show the evolution of the two-soliton input (each input soliton consists of carrying and weakly decaying modes). After decay of the accompanying mode we observe the characteristic dynamics of interacting in-phase solitons (i.e., of a breather) (cf.~\cite{two-soliton},~see also Ref.~\cite{suppl}).

{\it TM soliton --} In the spectrum of TM modes there can exist EPs~\cite{MK}. This makes the properties of TM modes very different as compared with TE-modes considered above.  Let the impedance $\eta^{TM}$ be chosen in the  vicinity of an EP, i.e., $\eta^{TM} = \eta^{EP} + r e^{i\Theta}$, where $r\ll 1$ (for a given statement $\eta^{EP}$ has a specific numerical value, but can be varied by modifying setting of the problem~\cite{suppl}). Although, strictly speaking the small amplitude expansion leading to Eqs.~(\ref{NLS1}) and (\ref{NLS2}) fails in the  {neighbourhood} of $\eta^{EP}$, we are interested exclusively in the phase behavior. Then, taking into account that at EP {\em two} eigenvalues coalesce, one can expand $Q\approx Q^{EP}+\nu e^{i\vartheta}$ where $\nu\sim\sqrt{r}\ll 1$ is a small parameter, while  $\vartheta=\Theta/2+\vartheta_0$, where $\vartheta_0=$const, is the phase which is changed by $\pi$ when $\eta^{TM}$ encircles the EP, i.e., when $\Theta$ is changed by $2\pi$.
Let the coalescing modes be of cosine type, i.e. $\phi(y)=\cos (Qy)$. Then in the leading order of the effective nonlinearity $g$ takes the form
\begin{eqnarray}
\label{nonlin_EP}
g\approx -  \chi e^{-i\vartheta} \frac{ \int_{-1}^{1} \cos^2(Q^{EP}y)|\cos(Q^{EP}y)|^2dy} {\nu \int_{-1}^{1} y \sin(2Q^{EP}y) dy }.
\end{eqnarray}  
Here we used the self-orthogonality of the eigenfunctions in the EP (see e.g.~\cite{Rotter}):   $\int_{-1}^{1} \cos^2(Q^{EP}y)dy=0$. Thus $\varphi_{g}=-\vartheta +$const and  the argument of $g$  changes by $\pi$
 when   $\eta^{TM}$ encircles $\eta^{EP}$.   According to the conditions (\ref{g_chi}) this means that the type of the effective nonlinearity changes (form focusing to defocusing or {\it vice versa}) independently of the material nonlinearity $\chi$.

If $Q$ is located away from the EP, the total phase $\varphi$ becomes nonlinearly dependent  on the rotation angle $\Theta$. This dependence in function of the ``distance" $r$ between $\eta^{TM}$ and the EP is illustrated in two central panels of Fig.~\ref{fig:two}. In the figure the change of the rotation angle $\Theta$ corresponds to the ``motion" along the curves in the direction indicated by arrows. The  striking situation of the opposite signs of the physical and effective nonlinearities is observed when the parameters ``move" from black discs (Re$\,\chi>0$, $g>0$) to the white discs (Re$\,\chi>0$, $g<0$). With the increase of $r$ the smaller rotation angle $\Theta$ is needed to achieve the total-phase change $\pi$. At the cladding impedance $\eta^{TM}$ corresponding to the black discs the effective nonlinearity is focusing and bright solitons can propagate in the system. An example is shown in the upper panels of Fig.~\ref{fig:two}.  We observe very robust evolution of the guided mode, even if at the input a weakly decaying mode is excited as well. Now the weakly decaying mode has the same parity as the guided one [shown by  red and blue squares in Fig.~\ref{fig:two}(a)] since both of them coalesce in the EP. The energy carried by both modes is concentrated near the absorbed boundaries. Unlike in the TE case, now the guided mode is not the fastest one: the largest positive propagation constant belongs to the decaying sine mode [the right triangle in Fig.~\ref{fig:two} (a)].

When the physical and effective nonlinearities are of different signs, in a waveguide with focusing nonlinearity there can propagate a stable dark soliton. In Fig.~\ref{fig:two} this is the situation corresponding to the white discs in panels (c) and (d). The stable evolution of a guided dark soliton excited at the input together with weakly decaying dark soliton, is illustrated in Fig.~\ref{fig:two}(f). Interestingly, while the structure of the modes remains similar to that of the bright soliton obtained for the same nonlinearity now the guided and weakly decaying modes are ``exchanged" [c.f. the insets and the location of red and blue squares in Figs.~\ref{fig:two}(a) and ~\ref{fig:two}(e)]. Similarly one can design a  waveguide with defocusing core nonlinearity supporting the propagation of bright TM polarized solitons. Finally, we mention the possibility of anomalous enhancement of the nonlinearity, which stems from the non-Hermitian nature of the system allowing the inner product $\langle\psi,\phi\rangle$ to be infinitely small which  leads to anomalously large effective nonlinearity $g$, seen  from Eqs.~(\ref{g}) and (\ref{nonlin_EP}) where $g\to\infty$ at $\nu\to 0$~\cite{suppl}.

 In conclusion, we reveal the key features of a dissipative waveguide with a nonlinear active core and absorbing boundaries which allow for the propagation of single-mode solitons and attenuate all other modes excited at the input. The type of the effective nonlinearity (focusing {\it vs} defocusing), as well as its absorbing or active characteristics are controlled by the boundary conditions. If the spectrum of the linear modes features EPs, the effective nonlinearity may acquire a sign opposite to the sign of the material nonlinearity of the core, which stems from the geometric phase acquired by the eigenfunctions when the impedance encircles the EP. In such situations the focusing (defocusing) Kerr nonlinearity can support propagation of dark (bright) solitons. The solitons reported are  structurally stable: the dependence on the waveguide parameters is continuous under the change of the parameters assuring the existence of the guided mode, while weak deviation of the parameters from the ideal guiding conditions results only in weak net dissipation or gain.
   An  important practical output, is that in the reported structures the sign of the effective nonlinearity can be changed {\em in situ} when the physical characteristics of the boundary are changed (by remote similarity with the atomic physics, where the change of the nonlinearity type is achieved by the Feshbach resonance).  Although we used the impedance boundary conditions, the reported effects are accessible with other types of absorbing boundaries and other types of dielectric filling. In particular, by using cladding with different impedances or made of metasurfaces~\cite{meta}, or birefringent filling one can control the position of the EP in the complex plane.

\smallskip

\begin{acknowledgments}
{\it Acknowledgment.} B.M.  was supported by the People Programme (Marie Curie Actions) of the European UnionÕs Seventh Framework Programme (FP7/2007-2013) under REA Grant No.~$[291734]$.\\
\end{acknowledgments}

\onecolumngrid
%\textwidth 7.in 

\medskip
\vspace{.25cm}
\begin{center}
{\textbf{\large Supplemental material}}
\end{center}

\medskip
%\medskip
\twocolumngrid
%%%%%%%%%% Merge with supplemental materials %%%%%%%%%%
%%%%%%%%%% Prefix a "S" to all equations, figures, tables and reset the counter %%%%%%%%%%
\setcounter{equation}{0}
\setcounter{figure}{0}
\setcounter{table}{0}
\makeatletter
\renewcommand{\theequation}{S\arabic{equation}}
\renewcommand{\thefigure}{S\arabic{figure}}
%\renewcommand{\bibnumfmt}[1]{[S#1]}
%\renewcommand{\citenumfont}[1]{S#1}
%%%%%%%%%% Prefix a "S" to all equations, figures, tables and reset the counter %%%%%%%%%%

\renewcommand{\theequation}{S\arabic{equation}}
\renewcommand{\thefigure}{S\arabic{figure}}
%\renewcommand{\bibnumfmt}[1]{[S#1]}
%\renewcommand{\citenumfont}[1]{S#1}

%\textheight{9cm}
%\renewcommand{\baselinestretch}{1.005}
\makeatother

\subsection{{\fontfamily{ptm}\selectfont Derivation of the  paraxial approximation Eqs. (3), (4)}}
For the sake of completeness, here we present a complete formal derivation of the paraxial approximation [Eqs. (3) and (4) of the main text] for the case of Robin boundary conditions. The derivation is given for the TE wave (for TM the derivation is similar).

Let us  start with 
\begin{eqnarray}
\label{Helmholtz1}
\nabla^2 E + \ell^2 k_0^2\epsilon E +\chi |E|^2 E =0.
\end{eqnarray}
and introduce the formal small parameter $\mu\ll 1$ defining the expansion for the field amplitude:
\begin{eqnarray}
\label{exp_field}
E=\mu E_0+\mu^2E_1+\mu^3 E_2+\cdots 
\end{eqnarray}
as well as the scaled variables $\{x_0,x_1,...\}$ and $\{z_0,z_1,...\}$ with $x_j=\mu^jx$ and    $z_j=\mu^j z$, which are treated as independent, so that
%\begin{subequations}
\begin{equation}
\label{exp_variab_x}
\frac{\partial}{\partial x}=\frac{\partial}{\partial x_0}+\mu \frac{\partial}{\partial x_1}+\cdots\,,
%\\
%\label{exp_variab_z}
\quad \frac{\partial}{\partial z}=\frac{\partial}{\partial z_0}+\mu \frac{\partial}{\partial z_1}+\cdots
\end{equation}
%\end{subequations} 
Substitution (\ref{exp_field}) and (\ref{exp_variab_x})
%and (\ref{exp_variab_z}) 
in Eq.~(\ref{Helmholtz1}) we obtain a series of equations at different orders of $\mu$
\begin{eqnarray}
\label{mu1}
\mu=1:\quad  & & {\mathcal L}E_0=0, \qquad {\mathcal L} \equiv \nabla_0^2+\ell^2 k_0^2\epsilon, 
\\
\label{mu2} 
\mu=2:\quad  & &{\mathcal L}E_1= 2\frac{\partial^2E_0}{\partial x_0\partial x_1}+2\frac{\partial^2E_0}{\partial z_0\partial z_1}, 
\\ 
\label{mu3}
\mu=3:\quad  & &{\mathcal L}E_2=   2\frac{\partial^2E_1}{\partial x_0\partial x_1}+2\frac{\partial^2E_1}{\partial z_0\partial z_1} +\frac{\partial^2 E_0}{\partial x_1^2}+ \frac{\partial^2 E_0 }{\partial z_1^2} 
\nonumber \\
& & \quad \quad + 2 \frac{\partial^2E_0}{\partial x_0\partial x_2}+2\frac{\partial^2E_0}{\partial z_0\partial z_2}   +\chi |E_0|^2 E_0,
\end{eqnarray}
where (notice that for the $y$-variables no scaling is needed)
$$\nabla_0 \equiv \left(\frac{\partial}{\partial x_0},\frac{\partial}{\partial y},\frac{\partial}{\partial z_0}\right).$$

Consider now the  {Sturm-Liouville} eigenvalue problem
\begin{eqnarray}
\phi_{yy}=-Q^2\phi,\quad {\phi(\pm 1)=\pm\eta^{TE} \phi_y(\pm 1)},
\end{eqnarray}
and its conjugate
\begin{eqnarray}
\psi_{yy}=-\tQ^2\psi, \quad {\psi(\pm 1)=\pm{(\eta^{TE})}^* \psi_y(\pm 1)}.
\end{eqnarray}
Let us define
\begin{eqnarray}
\label{q}
q=\left(\ell^2k_0^2\epsilon-Q^2\right)^{1/2}=q'+i\mu q'',
\end{eqnarray}
where $q',q''$ are real and the branch is {chosen} to ensure $q'\geq 0$. As a matter of fact (\ref{q}) reveals also the physical sense of the small parameter: it is a relation of the gain/loss coefficient to the propagation constant (in the original physical units). The spectra of the above {Sturm-Liouville} problems are discrete, and the eigenfunctions constitute the sets: $\{\phi_0,\phi_1,\phi_2...\}$ and $\{\psi_0,\psi_1,\psi_2...\}$, respectively. We can choose the numbering of the eigenfunction $\phi_n$ and the respective eigenvalues $Q_n$ such that $q_n''=\mbox{Im}(q_n)$ are ordered as 
\begin{eqnarray}
q_0''<q_1''<q_2''<\cdots
\end{eqnarray} 
Then the set for $\psi_n$ is determined by: $\psi_n=\phi_n^*$. Obviously, this ordering corresponds to higher modes undergoing stronger attenuation (or weaker amplification if the lowest $q_n$ are negative, this case however will not be considered here).

It is straightforward to ensure that
\begin{eqnarray}
\label{ortog}
\langle\psi_n,\phi_m \rangle=\int_{-1}^{1}\psi_n^*(y)\phi_m(y)dy=0\quad\mbox{if $m\neq n$}.\quad
\end{eqnarray}

Let us now assume that at the input, i.e. at $z=0$, only the two lowest modes, $\phi_{0}$ and $\phi_{1}$ are excited. Respectively we look for a solution of (\ref{Helmholtz1}) in the form where
\begin{eqnarray}
\label{E0}
E_0=A_0\phi_{0}(y)e^{iq_0' z_0-q_0''z_2}+A_1\phi_{1}(y)e^{iq_1' z_0-q_1''z_2},\quad\quad
\end{eqnarray}
where $A_0$ and $A_1$ are functions of only slow variables $x_1,\, x_2,...$ and $z_1,\, z_2,...$. The relation (\ref{q}) ensures, that the so defined $E_0$ solves (\ref{mu1}).   

Turning to the second order of the expansion. A general from of $E_1$, which at the input is zero, now reads (since we are considering the parameters out of the exceptional point, the sets  $\{\phi_0,\phi_1,\phi_2,...\}$ and $\{\psi_0,\psi_1,\psi_2,...\}$ constitute a biorthogonal basis):
\begin{equation}
\label{E1}
E_1=\sum_{m\neq 0} B_m^{(0)}\phi_m(y)e^{iq_0' z_0} +\sum_{m\neq 1} B_m^{(1)}\phi_m(y)e^{iq_1' z_0},
\end{equation}
 where $B_m^{(0)}$ and $B_m^{(1)}$ are functions on slow variables only.   Since  $E_1=0$ at $z=0$ and we are looking for a solution independent on $x_0$ one ensures that (\ref{mu2}) is satisfied by all $B_m^{(0,1)}=0$ and $\partial A_{0,1}/\partial z_1=0$. Thus $E_1\equiv 0$, and $A_{0,1}=A_{0,1}(x_1,z_2)$, i.e. depend on $x_1,\,x_2,...$ and $z_2, \, z_3,...$.

Turning now to the third order in $\mu$, we compute the right hand side of (\ref{mu3}) in the form  
\begin{eqnarray}
&&  {\mathcal L}E_2 = \phi_0 e^{iq_0' z_0-q_0''z_2} \left(2iq_0'\frac{\partial A_0}{\partial z_2} + \frac{\partial^2 A_0}{\partial x_1^2}\right) \nonumber \\ 
&&+\phi_1 e^{iq_1' z_0-q_1''z_2} \left(2iq_1'\frac{\partial A_1}{\partial z_2} + \frac{\partial^2 A_1}{\partial x_1^2}\right) \nonumber  \\ 
&& + \chi e^{iq_0' z_0}\phi_0 \left(e^{-2q_0''z_2}|\phi_0|^2|A_0|^2+2e^{-2q_1''z_2}|\phi_1|^2|A_1|^2\right)A_0
\nonumber \\ 
&& + \chi e^{iq_1' z_0} \phi_1\left(2e^{-2q_0''z_2}|\phi_0|^2|A_0|^2+e^{-2q_1''z_2}|\phi_1|^2|A_1|^2\right)A_1
\nonumber \\ 
&& + e^{i(2q_0'-q_1')z_0}e^{-(2q_0''+q_1'')z_2}\phi_{0}^2\phi_1^*A_0^2A_1^* 
\nonumber \\ 
&&+ e^{i(2q_1'-q_0')z_0}e^{-(2q_1''+q_0'')z_2}\phi_{1}^2\phi_0^*A_1^2A_0^*.\quad 
\end{eqnarray} 
 The solvability of this equation requires 
\begin{equation} 
\label{solvabil}
\langle \psi_0,{\mathcal L}E_2\rangle=\langle \psi_1,{\mathcal L}E_2\rangle=0.
\end{equation} 
Now considering the lowest mode with $q_0''=0$, and denoting {$q_0 = q, q_1 = \tq$},  $\phi_0=\phi$ and $\phi_1=\tilde{\phi}$, using the orthogonality relation, collecting terms having the same propagation constant, i.e. the ones $\sim e^{iq_0' z_0}$ and $\sim e^{iq_1' z_0}$, and letting $\mu=1$ {(this does not violate the assumption made at the beginning, provided we scale the slow dependencies properly, i.e. all having the same order)},  from  (\ref{solvabil}) we obtain the two nonlinear Schr\"odinger equations (3) and (4) from the main text.

\begin{figure}[t!]
	\includegraphics[width=.45\textwidth]{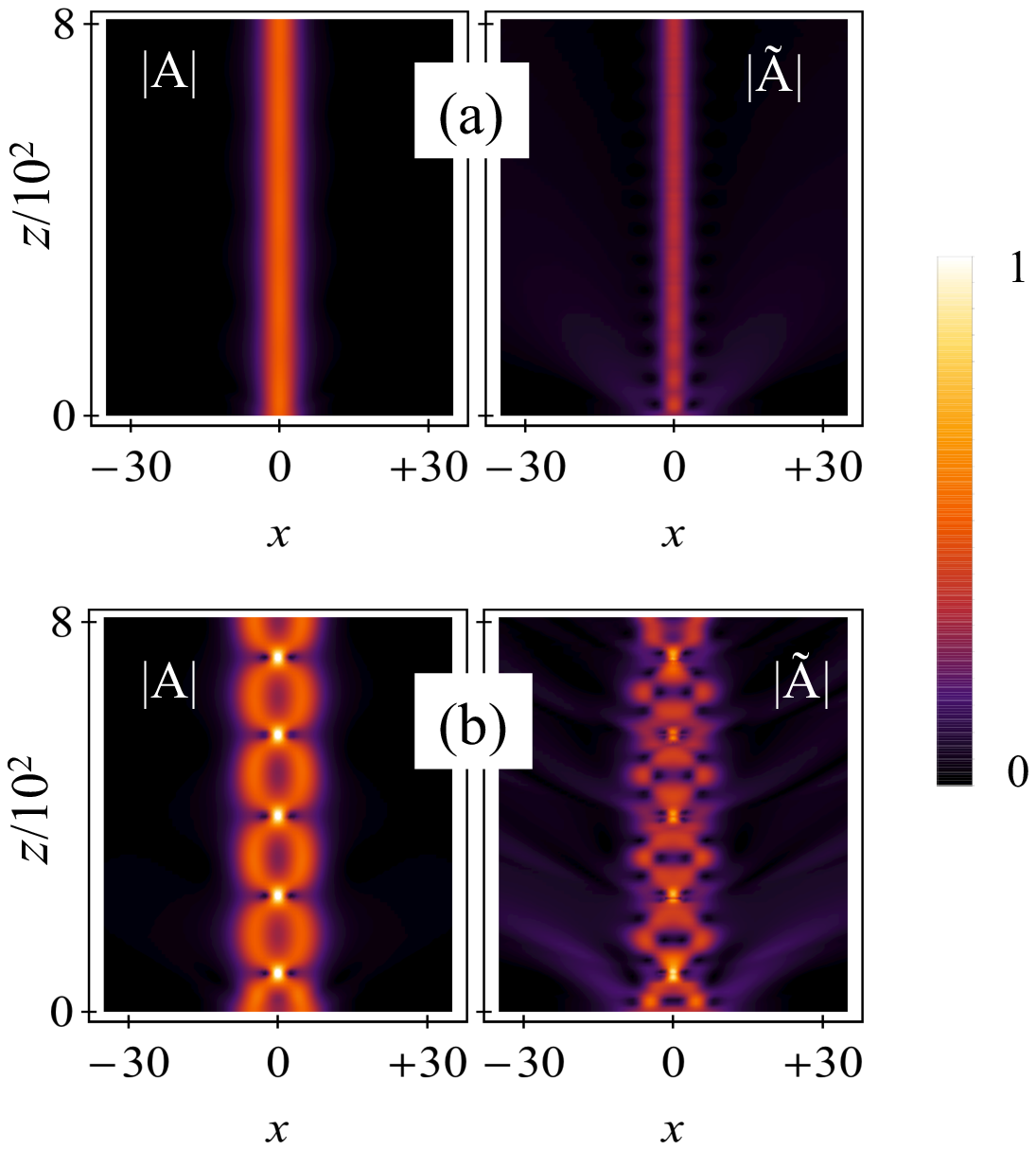}
	\caption{TE soliton mode intensities obtained after simulation of the coupled NLS equation (3) and (4) of the main text.  $A$ and $\tilde{A}$ shown in (a) corresponds to soliton 1(b) of main text; and those shown in (b) correspond to  soliton 1(c) of main text. Dynamics are done for the initial inputs :  (a) $A(x) = 2\tilde{A}(x) = 0.5 \sech(\sqrt{g/8} x)$, and  (b) $A(x) = 2 \tilde{A}(x) = 0.5 [\sech(\sqrt{g/8} (x+5)) + \sech(\sqrt{g/8} (x-5))]$. All the parameters are kept fixed as in Fig. 1 of the main text. }
	\label{NLS-solitons}
\end{figure}
\begin{figure}[t!]
	\includegraphics[width=.5\textwidth]{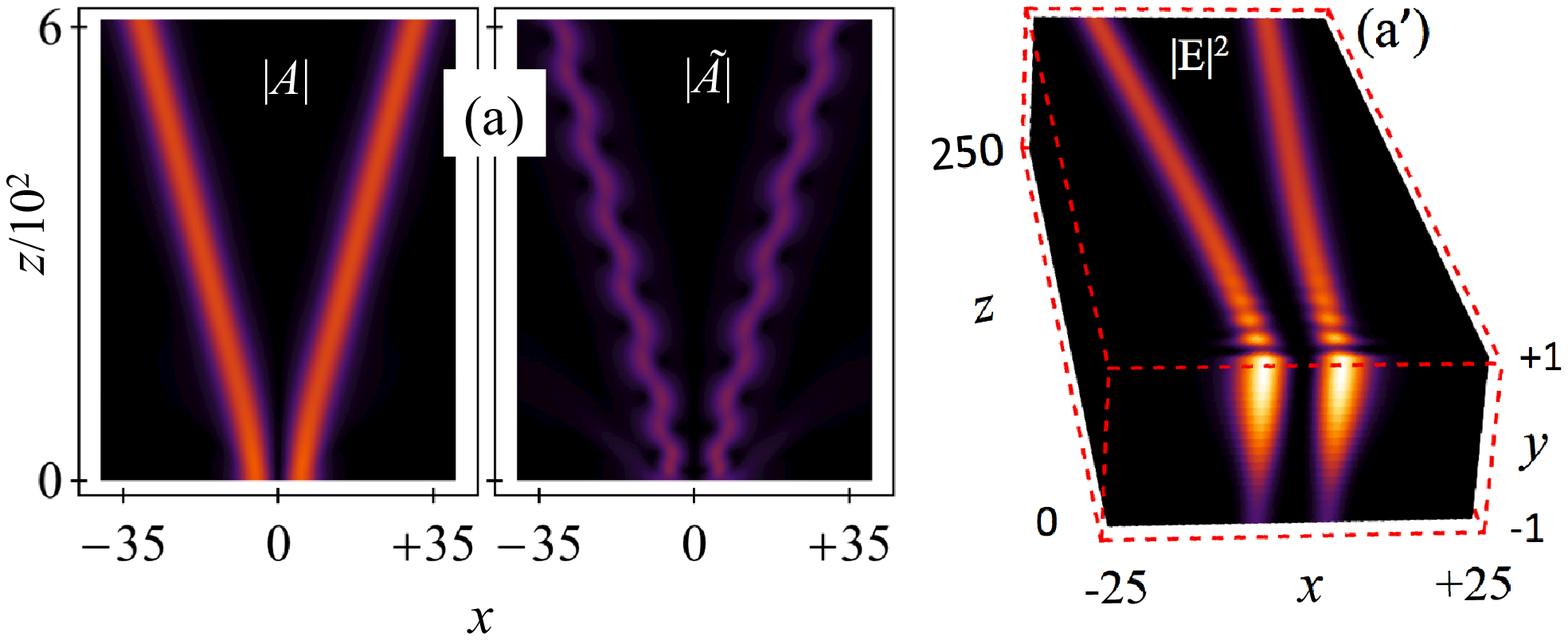}
	\caption{Simulation result for the TE soliton when two humped input solitons placed at $x=\pm5 $ and having relative initial phase $\pi$. In (a) mode amplitudes, and in (a$^\prime$) total electric field intensity are shown. All the parameters are kept fixed as in Fig. 1 of the main text.}
	\label{pi-solitons}
\end{figure}

Ensuring solvability of the Eq.~(\ref{mu3}), the small amplitude perturbations of the two lowest modes is searched [by analogy with (\ref{E1})] in the form
\begin{eqnarray}
\label{E2}
&&E_2=\sum_{m\neq 0}  C_m^{(0)}\phi_m(y)e^{iq_0' z_0} + \sum_{m\neq 1} C_m^{(1)}\phi_m(y)e^{iq_1' z_0}  +
\nonumber \\ 
&& \sum_{m} \left[D_m^{(0)}\phi_m(y)e^{i(2q_0'-q_1') z_0} + D_m^{(1)}\phi_m(y)e^{i(2q_1'-q_0')z_0}\right] \quad\quad~
\end{eqnarray}
where $C_m^{(0,1)}$ and $D_m^{(0,1)}$ are functions of slow variables. We are not interested here in the specific form of these coefficients, but a relevant fact is that by formally accounting for all harmonics in (\ref{E2}), the derivation assures that the approximation well describe the beam evolution of the lowest modes, provided the scaling of the applied beam is properly chosen.

A characteristic feature of the system (3), (4) from the main text, is that it includes decaying terms. In the main text we have shown that, in spite of this signature of the absorbing boundaries, the dynamics of the lowest modes is nearly integrable.  {This is also visible on the Fig.~\ref{NLS-solitons} where we show the evolutions of the  the mode intensities, $A(x,z)$ and $\tilde{A}(x,z)$, corresponding to Figure 1 of main text.} As an additional confirmation of this  fact, in Fig.~\ref{pi-solitons} we illustrate "repulsion" of two beams having identical intensities and the $\pi$-shifted phases. The observed dynamics closely resemble the dynamics of two out-of-phase solitons of the NLS equation. 

\begin{figure}[t!]
	\includegraphics[width=.42\textwidth]{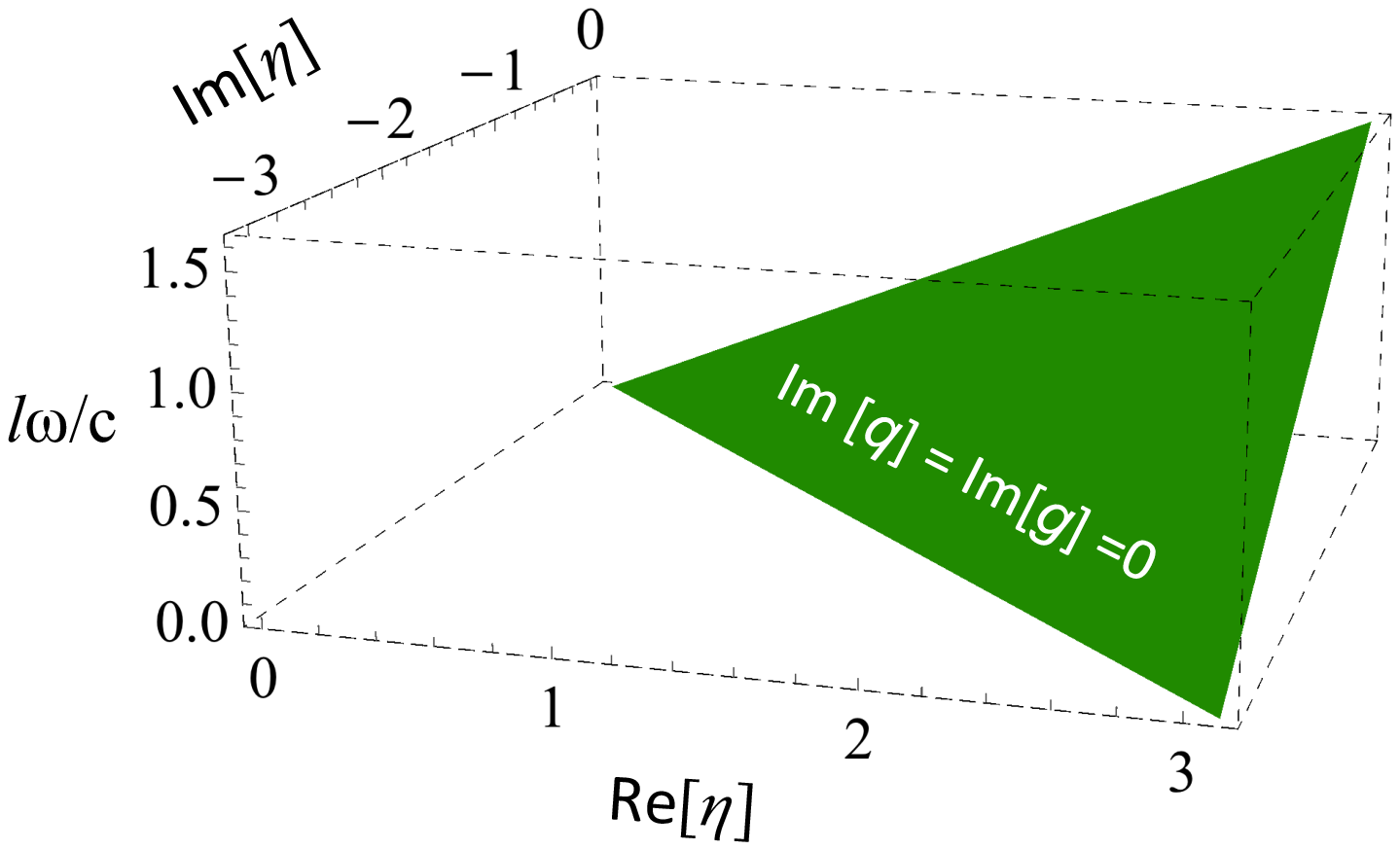}
	\caption{Existence region for non-diffracting (Im$[g] = 0$) and non-vanishing (Im$[q] = 0$) soliton which was elaborated in the figure 1 of main text. Other  parameters for this plot was kept fixed as in figure 1 of main text.}\label{stability-region}
\end{figure}

 Furthermore, it is important to mention here that the soliton shown in Figure 1 of the main text is point like in the effective parameter space. However, as we show below (in Fig.~\ref{stability-region}), the corresponding soliton persists for a wide range of physical parameters e.g. actual impedance $\eta$, $\ell$, and $\omega$, which are possible to control in experiments. Moreover, if admit that the condition for the main mode are not perfectly conducting, but result in weak absorption or gain, i.e. in a complex $q_0$, the phenomenon is still observable after the respective loss or gain described by $e^{-q_0''z}$ is scaled out. In this sense the phenomena reported in the main text are structurally stable.

\begin{figure}[h!]
\includegraphics[width=.5\textwidth,height=4.75cm]{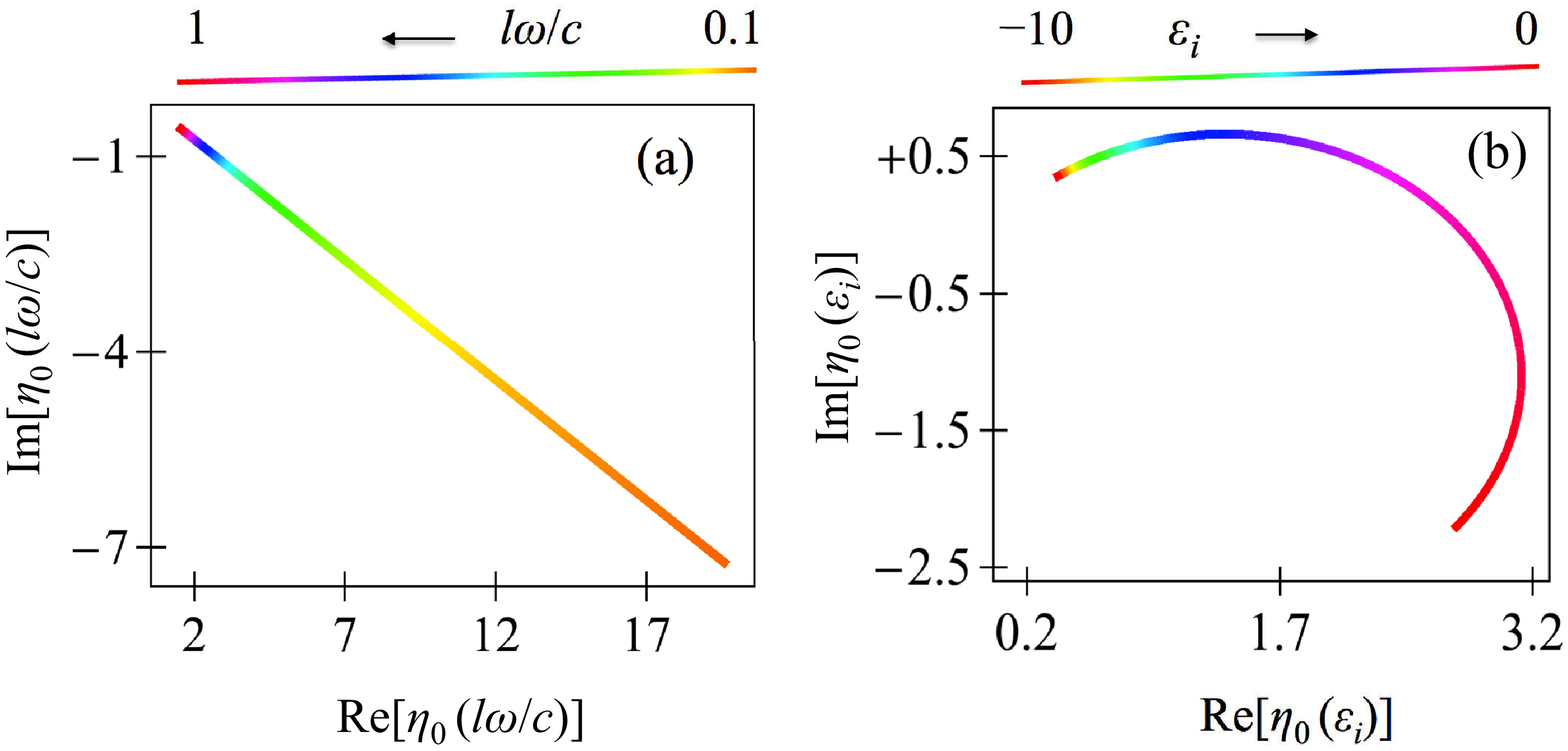}
\caption{Trajectory of the EP, $\eta_0$, in the complex $\eta$ plane with respect to the variation (range of variations are shown in the color bars placed above the respective figure) of (a) $\ell \omega/c$ when $\epsilon = (1.5 - 0.5~i)$ is kept fixed; and (b) gain distribution $\epsilon_i$ of the core, when $\epsilon_r=1.5$, $\omega = c/2\ell$.}\label{fig-suppliment-3}
\end{figure}

\subsection{{\fontfamily{ptm}\selectfont Control of exceptional point by the system parameters}}
As reported in Ref.~\cite{MK}, the system has infinitely many exceptional points which are defined in the complex effective impedance parameter,$\eta^{TM}$, plane. The effective impedance is a composite parameter consisting of all the physical parameters: $\eta^{TM} = i \eta \omega \epsilon \ell/c$. When expressed in terms of physical impedance $\eta$, the EP can be controlled by changing waveguide transverse dimension, $\ell$, or the frequency, $\omega$, of the incident field, or the dielectric constant, $\epsilon$, of the core. Here in Fig.~\ref{fig-suppliment-3}, we show the movement of the EP (for which $\eta^{TM}  \approx 1.65061 + 2.05998i$) in the complex $\eta$ plane for some variation of other parameters. In the figure we denote $\eta_0$ as the EP corresponding to the the parameter $\eta$.

A highly promising way of manipulating the parameters of the exceptional point is the use birefringent feeling, changing the relations between the ``transverse", $Q$, and the ``forward", $q$, propagation constants for the mode, which in the meantime add additional control parameters. This possibility requires more extending study. 

\begin{figure}[h!]
\includegraphics[width=0.43\textwidth,height=5cm]{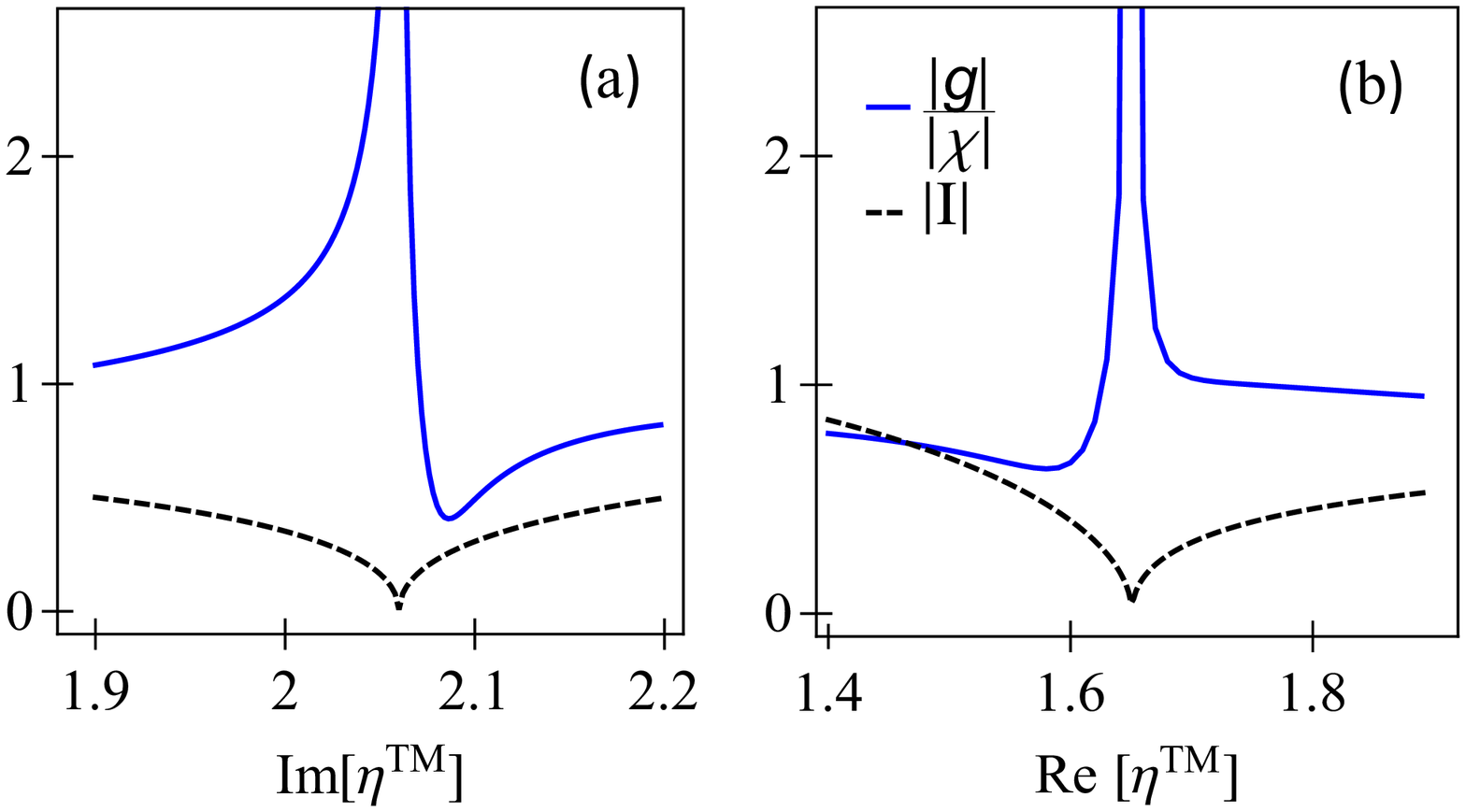} 
\caption{(a) Absolute value of $g$ (in the unit of $|\chi|$) {\it vs} Im[$\eta^{TM}$] when Re[$\eta^{TM}] =$ Re$[\eta^{EP}] =1.65$ is fixed; (b) absolute value of $g$ {\it vs} Re[$\eta^{TM}$] when Im[$\eta^{TM}] =$ Im$[\eta^{EP}] =2.06$ is fixed. The associated self-overlap integrals are shown black lines.}\label{fig-suppliment-1}
\end{figure}
\begin{figure}[t!]
\centering
\includegraphics[width=.48\textwidth,height=5cm]{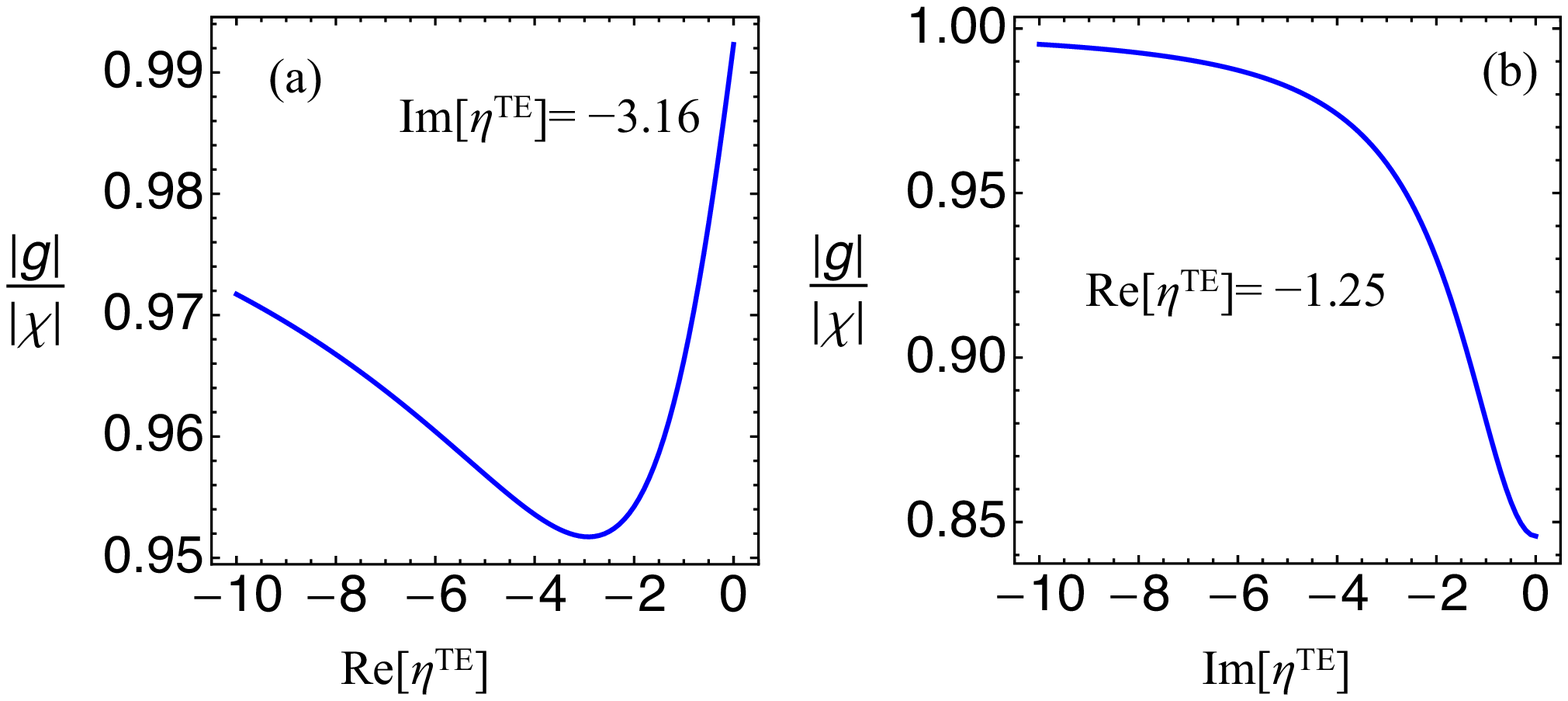}
\caption{(a) Absolute value of $g$ {\it vs} Re[$\eta^{TE}$] when Im[$\eta^{TE}] = -3.16$ is fixed; (b) absolute value of $g$ {\it vs} Im[$\eta^{TE}$] when Re[$\eta^{TE}] =-1.25$ is fixed.}\label{fig-suppliment-2}
\end{figure}

\subsection{{\fontfamily{ptm}\selectfont Anomalous enhancement of effective-nonlinearity near EP}}
At an EP, the eigenfunctions of the coalescent mode is self-orthogonal, i.e. $I = \langle\psi, \phi\rangle = \int_{-1}^1 dy~ \phi^2(y)\to 0$ when $\phi(y)$ approaches the the EP. This in turn implies that the effective nonlinearity $g$ [obtained in eq.(5) of main text] undergoes anomalous enhancement, because vanishing denominator, when the mode approaches an EP. In Fig.~\ref{fig-suppliment-1}, we show the numerical evidence of this enhancement, and associated self-overlap integrals when the impedance parameter varies near an EP. 

On the other hand, for the TE modes the changes of effective nonlinearity is shown in figure \ref{fig-suppliment-2}, 
which shows that both enhancement and decrement of effective nonlinearity are also possible.

\onecolumngrid

\end{document}